\documentclass[12pt]{article}

\usepackage{times}
\usepackage{geometry}
\usepackage[utf8]{inputenc}
\usepackage{enumitem,amssymb}
\usepackage{ragged2e}
\usepackage[numbers]{natbib}
\usepackage{wrapfig}
\usepackage{multicol}
\usepackage[citecolor=blue, colorlinks=True]{hyperref}
\usepackage{graphicx}
\newlist{thematic}{itemize}{8}
\setlist[thematic]{label=$\square$}
\usepackage{pifont}
\usepackage{color}
\usepackage{xcolor}
\newcommand{\mjup}{\,M_{Jup}}

\def\ref@jnl#1{{\jnl@style#1}}

\def\aj{\ref@jnl{AJ}}                   
\def\actaa{\ref@jnl{Acta Astron.}}      
\def\araa{\ref@jnl{ARA\&A}}             
\def\apj{\ref@jnl{ApJ}}                 
\def\apjl{\ref@jnl{ApJL}}                
\def\apjs{\ref@jnl{ApJS}}               
\def\ao{\ref@jnl{Appl.~Opt.}}           
\def\apss{\ref@jnl{Ap\&SS}}             
\def\aap{\ref@jnl{A\&A}}                
\def\aapr{\ref@jnl{A\&A~Rev.}}          
\def\aaps{\ref@jnl{A\&AS}}              
\def\azh{\ref@jnl{AZh}}                 
\def\baas{\ref@jnl{BAAS}}               
\def\bac{\ref@jnl{Bull. astr. Inst. Czechosl.}}
\def\caa{\ref@jnl{Chinese Astron. Astrophys.}}
\def\cjaa{\ref@jnl{Chinese J. Astron. Astrophys.}}
\def\icarus{\ref@jnl{Icarus}}           
\def\jcap{\ref@jnl{J. Cosmology Astropart. Phys.}}
\def\jrasc{\ref@jnl{JRASC}}             
\def\memras{\ref@jnl{MmRAS}}            
\def\mnras{\ref@jnl{MNRAS}}             
\def\na{\ref@jnl{New A}}                
\def\nar{\ref@jnl{New A Rev.}}          
\def\pra{\ref@jnl{Phys.~Rev.~A}}        
\def\prb{\ref@jnl{Phys.~Rev.~B}}        
\def\prc{\ref@jnl{Phys.~Rev.~C}}        
\def\prd{\ref@jnl{Phys.~Rev.~D}}        
\def\pre{\ref@jnl{Phys.~Rev.~E}}        
\def\prl{\ref@jnl{Phys.~Rev.~Lett.}}    
\def\pasa{\ref@jnl{PASA}}               
\def\pasp{\ref@jnl{PASP}}               
\def\pasj{\ref@jnl{PASJ}}               
\def\rmxaa{\ref@jnl{Rev. Mexicana Astron. Astrofis.}}%
\def\qjras{\ref@jnl{QJRAS}}             
\def\skytel{\ref@jnl{S\&T}}             
\def\solphys{\ref@jnl{Sol.~Phys.}}      
\def\sovast{\ref@jnl{Soviet~Ast.}}      
\def\ssr{\ref@jnl{Space~Sci.~Rev.}}     
\def\zap{\ref@jnl{ZAp}}                 
\def\nat{\ref@jnl{Nature}}              
\def\iaucirc{\ref@jnl{IAU~Circ.}}       
\def\aplett{\ref@jnl{Astrophys.~Lett.}} 
\def\apspr{\ref@jnl{Astrophys.~Space~Phys.~Res.}}
\def\bain{\ref@jnl{Bull.~Astron.~Inst.~Netherlands}} 
\def\fcp{\ref@jnl{Fund.~Cosmic~Phys.}}  
\def\gca{\ref@jnl{Geochim.~Cosmochim.~Acta}}   
\def\grl{\ref@jnl{Geophys.~Res.~Lett.}} 
\def\jcp{\ref@jnl{J.~Chem.~Phys.}}      
\def\jgr{\ref@jnl{J.~Geophys.~Res.}}    
\def\jqsrt{\ref@jnl{J.~Quant.~Spec.~Radiat.~Transf.}}
\def\memsai{\ref@jnl{Mem.~Soc.~Astron.~Italiana}}
\def\nphysa{\ref@jnl{Nucl.~Phys.~A}}   
\def\physrep{\ref@jnl{Phys.~Rep.}}   
\def\physscr{\ref@jnl{Phys.~Scr}}   
\def\planss{\ref@jnl{Planet.~Space~Sci.}}   
\def\procspie{\ref@jnl{Proc.~SPIE}}   

\begin{document}
\setlength{\textfloatsep}{8pt plus 1.0pt minus 2.0pt}
\raggedright
\huge
Astro2020 Science White Paper \linebreak

Substellar Multiplicity Throughout the Ages
\normalsize

\noindent \textbf{Thematic Areas:} \hspace*{60pt} \\
$\square$ Planetary Systems \hspace*{10pt} 
$\blacksquare$ Star and Planet Formation \hspace*{20pt}\linebreak
$\square$ Formation and Evolution of Compact Objects 
$\square$ Cosmology and Fundamental Physics\linebreak
$\blacksquare$ Stars and Stellar Evolution
$\square$ Resolved Stellar Populations and their Environments\linebreak
$\square$ Galaxy Evolution 
$\square$ Multi-Messenger Astronomy and Astrophysics \linebreak
  
\textbf{Principal Author:}

Name:	Daniella C. Bardalez Gagliuffi
 \linebreak						
Institution:  American Museum of Natural History
 \linebreak
Email: dbardalezgagliuffi@amnh.org
 \linebreak
Phone:  212-496-3444
 \linebreak
 
\textbf{Co-authors:} (names and institutions)
\linebreak
Kimberly Ward-Duong, Amherst College, Amherst, MA\\
Jacqueline Faherty, AMNH, New York, NY\\
Alex Greenbaum, U. Michigan, MI\\
Federico Marocco, JPL, Pasadena, CA\\
Adam Burgasser, UC San Diego, La Jolla, CA\\
Matthew Bate, U. Exeter, Exeter, UK\\
Trent Dupuy, Gemini Observatory, Hilo, HI\\
Christopher Gelino, IPAC, Pasadena, CA\\
Johannes Sahlmann, STScI, Baltimore, MD\\
Frantz Martinache, Observatoire de la C\^{o}te d'Azur, Nice, France\\
Michael Meyer, U. Michigan, MI\\
Quinn Konopacky, UC San Diego, La Jolla, CA\\
Denise Stephens, Brigham Young University, Provo, UT\\
\medskip
\textbf{Abstract  (optional):}
\justify

Substellar multiplicity is a key outcome of the formation process. The biggest challenge for the next decade will be to distinguish between the formation history, environmental conditions, and dynamical evolution leading to the least massive brown dwarfs and the most massive planets at the tail ends of their mass functions. In this white paper, we advocate for a comprehensive characterization of both the statistical distributions of the population of ultracool dwarf multiple systems and the fundamental properties of their individual components as a function of age. A space-based precision astrometry mission in near-infrared wavelengths would provide the necessary measurements to identify and characterize age-calibrated populations of multiple systems.


\pagebreak


\section{Multiplicity as a probe of substellar formation and evolution}

Brown dwarfs are substellar objects lacking sufficient mass to sustain hydrogen fusion in their cores, becoming colder and fainter with age~\citep{1963ApJ...137.1121K}. Brown dwarfs span the mass range between the lowest-mass stars and the most massive planets ($M_{BD}\sim13-80\mjup$), although the low-mass bound is arbitrarily set by the nominal cut-off for deuterium burning~\citep{2011ApJ...727...57S}.  Similar to stars, brown dwarfs can be formed in binary or higher order multiple systems, which are coeval and cospatial laboratories with shared composition, making them excellent probes of their formation and evolution processes. Many key questions regarding substellar formation and evolution remain, and can be answered with measurements of brown dwarf multiplicity as a function of mass, age, composition and environment, possibly revisiting the definitions of brown dwarfs and giant planets.

Although brown dwarfs appear to be the lowest-mass products of star formation, the specific processes leading to their formation are poorly understood. In the early 2000s, four scenarios were proposed to explain their formation, all of which create conditions to limit the available mass for a forming proto-star to accrete. These scenarios include: turbulent fragmentation in high density  layers~\citep{2002ApJ...576..870P}, fragmentation of a proto-stellar disk~\citep{2009MNRAS.400.1563S}, ejection from a stellar nursery~\citep{2001AJ....122..432R}, and photo-erosion through stellar winds from a nearby OB star~\citep{2004AandA...427..299W}, and have been explored through numerical simulations with various predictions. N-body simulations of disintegrating triple systems predict a brown dwarf binary fraction up to 43\%~\citep{2015AJ....149..145R}. Disk fragmentation models predict an ultracool ($M\le0.1\,M_{\odot}$) binary fraction of $\sim$16\%, including hierarchical triples with a main sequence host and ejected tight binary systems~\citep{2009MNRAS.392..413S}. Radiation hydrodynamical simulations find multiplicity fractions that continually decrease with primary mass throughout the low-mass star and substellar regimes, leading to $\sim 8$\% for $0.018 - 0.100 M_{\odot}$ primaries~\citep{2012MNRAS.419.3115B}. More advanced calculations find $\approx 15$\% in the mass range of $0.07 - 0.10 M_{\odot}$, falling to $\approx 4$\% in the mass range of $0.03 - 0.07 M_{\odot}$ with little evidence for a metallicity dependence~\citep{2019MNRAS.484.2341B}. Observations of the binary fraction in star-forming regions and the field are limited by small samples or low resolution, and the results depend on the technique used for binary identification (see Section 1.1).

At the low-mass end of the brown dwarf regime, binary formation may start to look like planet formation with a brown dwarf host. A few systems challenge our understanding of the distinctions between stellar and planet formation channels, like 2MASS~J1207$-$3932, composed of a 5\,$M_{Jup}$ planetary-mass object orbiting 44\,AU away from a 33\,$M_{Jup}$ brown dwarf~\citep{2015ApJ...810..158F}, also identified as the first directly-imaged ``planet''~\citep{2004AandA...425L..29C}. Recent studies suggest a transition from a stellar-like gravitational instability formation pathway to a planetary-like core accretion mechanism (where planetesimals grow from individual particles until they are massive enough to retain a gaseous atmosphere;~\citep{1996Icar..124...62P}) around $4-10\mjup$~\citep{2018ApJ...853...37S}, supported by the observation that core accretion has difficulties forming planets more massive than $\sim5\mjup$~\citep{2018haex.bookE.143M,2007ApJ...662.1282M}.

As hybrids of stars and planets, do brown dwarfs also share formation mechanisms with both? Is there a clear mass cut-off in the products generated by each formation channel? The current challenge is to explore the universality of brown dwarf formation in different environments, and one approach is through measuring multiplicity. More complete measurements of the occurrence of multiple systems and the distribution of their separations, relative masses and eccentricities, can help constrain formation theories. While measurements for nearby, field-age systems are more accessible than measurements of systems in coeval clusters with our current instrumentation, it is crucial to measure multiplicity statistics and fundamental parameters of the individual components across different ages and environments to trace evolutionary processes.



\subsection{What is the binary fraction of the lowest-mass brown dwarfs?}

Stellar multiplicity is characteristic of stellar spectral type, with multiplicity frequency decreasing with primary mass: $>60-70$\% in B-stars~\citep{2012Sci...337..444S, 1990ApJS...74..551A}, likely $>50$\% for A-stars~\citep{2013ARAandA..51..269D}, $\sim$44\% in FGK dwarfs~\citep{2010ApJS..190....1R}, and $27\pm5\%$ for M dwarfs~\citep{1997AJ....113.2246R}. In the early 2000s, the \textit{Hubble Space Telescope} and ground-based adaptive optics (AO) systems enabled multiplicity studies across and beyond the stellar/substellar boundary, leading to resolved binary fractions of 10\% for $>$M7 dwarfs~\citep{2003AJ....126.1526B}, $15\pm5$\% for late-M and L dwarfs~\citep{2003ApJ...587..407C,2003AJ....125.3302G,2005ApJ...621.1023S}, and $9^{+15}_{-4}$\% for T dwarfs~\citep{2003ApJ...586..512B,2006ApJS..166..585B}. In the last decade, volume-limited samples of 15\,pc and 25\,pc provided multiplicity fractions of 29$\pm$3\% for K7$-$M6 dwarfs~\citep{2015MNRAS.449.2618W} and 27$\pm$1\% for M0$-$M9 dwarfs~\citep{2019arXiv190106364W}. Ongoing studies aim to refine the multiplicity fraction across the  M/L (Bardalez Gagliuffi et al., in prep.) and L/T spectral classes (Best et al., in prep.). In the last 20 years, high resolution imaging surveys found the first companions to mid- to late-T dwarf primaries~(e.g. \citep{2003ApJ...586..512B,2011AJ....142...57G,2016ApJ...833...96L}), setting constraints on binary fraction; e.g. $<16\%$~\citep{2014AJ....148..129A} and $8\pm6\%$~\citep{2018MNRAS.479.2702F}. No companions have been found yet to Y dwarf primaries~\citep{2016ApJ...819...17O}. Due to the intrinsic faintness and scarcity of the coldest brown dwarfs, these studies evaluated small samples (26 mid- to late-T dwarfs,~\citep{2014AJ....148..129A}; 12 T8$-$Y0 dwarfs,~\citep{2018MNRAS.479.2702F}; and 5 Y dwarfs,~\citep{2016ApJ...819...17O}), leaving room for discovery and characterization with upcoming instrumentation.

Understanding dynamical evolution requires comparison of binary fractions at different ages. Multiplicity fractions across star-forming regions and intermediate-age clusters are statistically equivalent to those of older populations, with larger uncertainties and a higher proportion of low mass ratio systems, indicating observational bias in field studies, e.g., $10^{+18}_{-8}\%$ for 2~Myr in Chamaeleon~\citep{2008AandA...492..545J}, 33$\pm$17\% for 5-10~Myr in Upper Scorpius~\citep{2005ApJ...633..452K}), a limit of $<11$\% for the lowest-mass T-dwarf precursor population of 11 objects in the 125\,Myr Pleiades cluster~\citep{2015ApJ...804...65G}, and $<19\%$ in the 600\,Myr Hyades cluster~\citep{2013AandA...555A.137D}. While at least two substellar binaries have been identified in the 12\,Myr $\beta$ Pictoris association~(2MASS J0249$-$0557~\citep{2018AJ....156...57D} and 2MASS J1207$-$3932~\citep{2004AandA...425L..29C}), there has been no systematic search for binaries in young moving groups.  Nevertheless, these studies are underscored by the ambiguity that remains from small samples and small number statistics.

Recent studies are pushing the limits of low-mass companions as well as the definitions of low-mass binaries~\citep{2011ApJ...730...39B,2019arXiv190302332F}.  In addition to 2MASS J1207$-$3932 \citep{2004AandA...425L..29C}, only a handful of these systems are known (e.g. 2MASS J0441+2301Bb, a $10\pm2\mjup$ object at 15\,AU from a $19\pm3\mjup$ brown dwarf, in a quadruple system in Taurus;~\citep{2014ApJ...788...40T}). Systems straddling the deuterium-burning limit are critical to constrain brown dwarf and planetary formation channels. One of the main objectives for the next decade will be constraining binary fraction of the coldest and lowest-mass brown dwarfs with \textit{JWST} and extremely large telescopes (ELTs), while determining the brown dwarf or planetary nature of these very low-mass companions. Understanding the underlying mass ratio distribution and compositional differences between brown dwarfs and giant planets is key to trace observations to formation mechanisms. 



Perhaps the most stringent constraints on both evolution and formation are provided by wide binaries, hierarchical triple systems, higher-order multiple systems, and ultracool companions to main sequence stars (i.e. ``brown dwarf desert'' systems). Wide binaries are difficult to form in situ\footnote{However, see~\citep{2004ApJ...614..398L} for a description of 2MASS J1101$-$7732, a 240\,AU, M7+M8 binary in Chamaeleon.}, and likely to disrupt over time due to low binding energies~\citep{2016MNRAS.459.4499E}. Many wide binaries could in fact be hierarchical triple systems (see~\citep{2010ApJ...720.1727L} for M1$-$M5 dwarf primaries), which would help explain their persistence to field ages. However, only 5 triple systems with total mass $M_{tot}\leq0.3\,M_{\odot}$ have been confirmed~\citep{2000MNRAS.311..385G,2006ApJ...645L.153P,2008AandA...484..429S, 2012ApJ...757..110B,2013ApJ...778...36R}, including one in the 12\,Myr $\beta$ Pictoris young moving group~\citep{2018AJ....156...57D}. This particular configuration could result from migration of the lowest-mass component by Kozai-Lidov oscillations~\citep{1961AJ.....66..132K,1962PandSS....9..719L}, or the wide binary pair may be sufficient to organize into a hierarchical triple with a planetary-mass companion~\citep{2019arXiv190302332F,2012ApJ...754L..36N,2016ApJ...827....8N}. 

At any separation, brown dwarf companions to main sequence stars and white dwarfs are priceless evolutionary benchmarks since the primary can provide an independent measure of age~\citep{2011MNRAS.410..705D}. However, the occurrence of these systems is low, particularly at small separations (i.e., $<5\,$AU), a paucity known as the ``brown dwarf desert''~\citep{2006ApJ...640.1051G,2011ApJ...731....8K,2006ApJ...640.1051G,2019arXiv190302332F}. At wide separations, we know of 4 planetary-mass brown dwarf companions to young main sequence stars~\citep{2014ApJ...787....5N,2011AJ....141..119K,2010MNRAS.405.1140G,2016MNRAS.457.3191D}.
 
 


\subsection{Detection techniques sample different ranges of binary separations}

High resolution imaging studies lead to the discovery of a peak in the separation distribution of systems with an M7 primary or later-type at $4-7$\,AU, with an apparent shortage of wide systems (93\% at $<20$\,AU) and tightly-separated ones~\citep{2007ApJ...668..492A,2007prpl.conf..427B}. The location of this peak follows a decreasing trend in typical separation from binaries with higher mass primaries, possibly indicating a preferred formation scale~\citep{2014MNRAS.442.3722P}. Direct imaging has identified over 85\% of the known substellar binaries~\citep{2015AJ....150..163B}, though it might be biased against tightly-separated systems. Due to the angular resolution of ground-based adaptive optics\footnote{Most substellar objects are fainter ($I>15\,$mag) than the limit of current-generation extreme-AO systems~\citep{2014PNAS..11112661M}.}, inner working angles for substellar sources roughly coincide with the peak in the separation distribution, suggesting an observational bias. 

Thanks to~\textit{Gaia}, it is now possible to identify binary systems in bulk by over luminosity in color-magnitude diagrams~\citep{2018arXiv180908244R}, while this technique was previously restricted to observationally-expensive parallax programs~\citep{2009AJ....137....1F,2013AandA...560A..52M}. Additionally, the 5-parameter kinematic solutions permits large, systematic studies to identify wide, common proper motion companions~\citep{2017AJ....153..257O,2018RNAAS...2c.137K}. 

More resource-intensive binary detection techniques, like radial velocity (RV) and astrometric variability, have detected closely-separated binary systems otherwise missed by imaging in small samples, also yielding lower binary fractions (e.g. $2.5^{+8.6}_{-1.6}$ for $<1\,$AU with RV~\citep{2010ApJ...723..684B}, $\sim5\%$~\citep{2015AandA...577A..15S}). High-precision astrometry and radial velocity follow-up of $\sim20$ brown dwarf binaries have allowed full orbit and dynamical masses determination, and the first few tests of substellar evolutionary models (see white paper by T. Dupuy), some of which have revealed discrepancies between measured and predicted brown dwarf masses~\citep{2010ApJ...711.1087K,2013AandA...556A.133S,2017ApJS..231...15D}.


Both radii and dynamical mass measurements can be directly obtained from eclipsing binaries.  However, only one such system is known, 2MASS J0535$-$0546~\citep{2006Natur.440..311S} in the 1\,Myr Orion Nebula Cluster, despite most brown dwarf binaries having small separations ($<10\,$AU). Increasing the number of substellar eclipsing binaries would require long-term monitoring, akin to \textit{Kepler} or \textit{TESS} observations, both of which operate in the optical where most brown dwarfs are too faint.

Direct individual mass measurements can also be achieved through microlensing. About a dozen substellar binary systems have been identified through this technique~\citep{2008ApJ...684..663B,2010ApJ...723..797H,2013ApJ...768..129C,2013ApJ...778...38H,2017AJ....154..223H,2017ApJ...843...59H, 2017AandA...604A.103P,2017ApJ...840L...3S,2018ApJ...858..107A}, most of them at sub-AU separations and kiloparsec distances, including some straddling the deuterium burning limit. These systems are too faint and too far away to be followed-up with any existing or planned facility. In the near future,~\textit{WFIRST} will return plenty of low-mass microlensing binaries for statistical treatment but no individual characterization. 

Blended-light spectral binaries provide a useful screening technique that can identify short-separation systems with components of different spectral morphologies from a single low resolution, spectrum~\citep{2010ApJ...710.1142B,2011ASPC..447..177H,2014ApJ...794..143B}. These alternative techniques are starting to fill the gap of the smallest orbital separations and avoid the bias against extreme mass ratio systems. These systems must also be followed up, as the distributions of orbital parameters like the separation, mass ratio, and eccentricity provide detailed tests of formation simulations. 





\begin{figure}
\begin{center}
\vskip -0.2in
\includegraphics[scale=0.5]{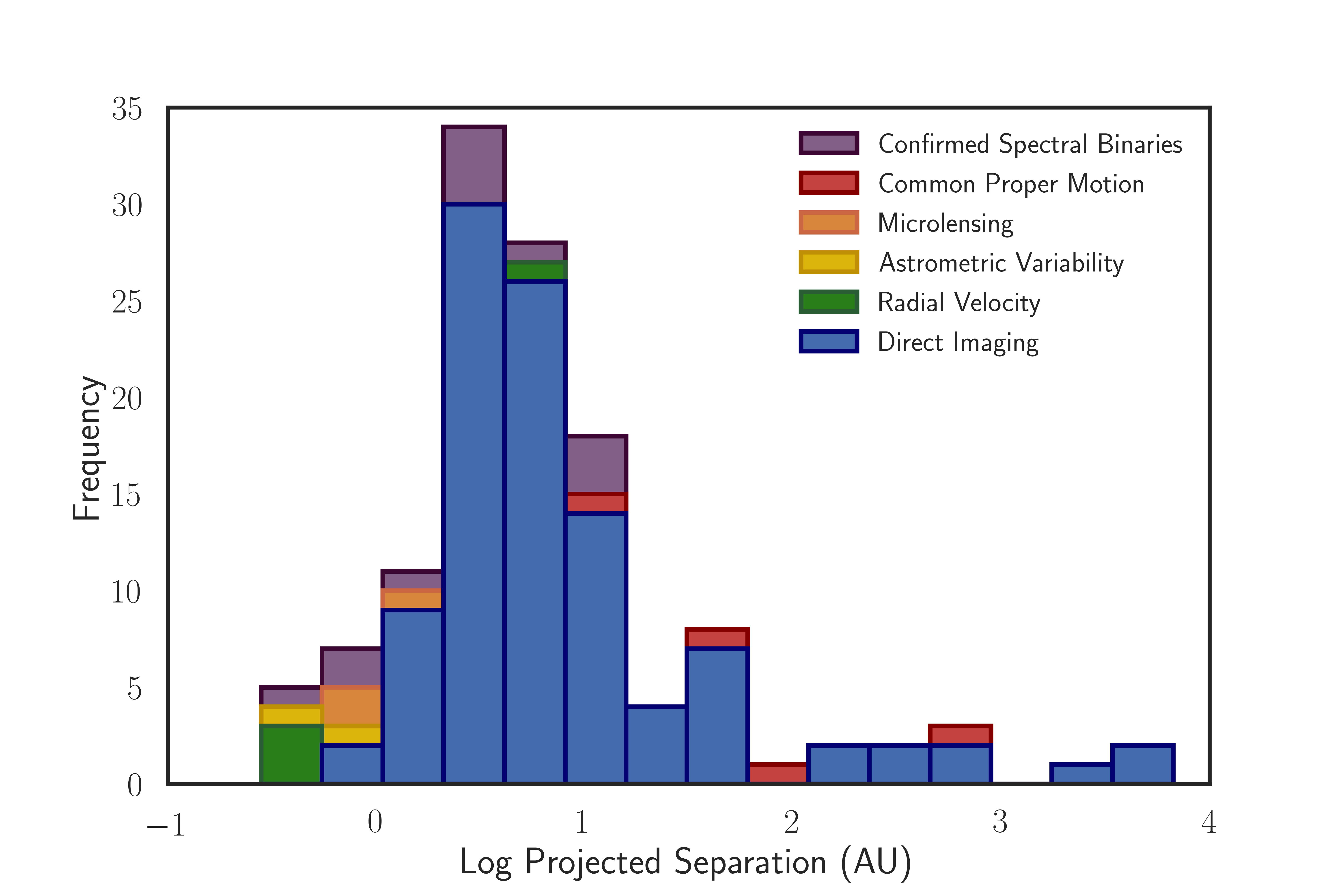}
\caption{\small Projected separation distribution of ultracool binaries discovered up to 2014, showing the separations sampled by each technique in practice.}
\end{center}
\end{figure}

\section{Opportunities}

In the last two decades, we have significantly advanced our understanding of multiplicity across the hydrogen-burning limit. In this upcoming decade, we will do the same across the deuterium-burning limit, and explore the distinctions of star and planet formation mechanisms. Constraining the multiplicity fractions and the statistical distributions (i.e. mass ratio, separation, eccentricities) of the brown dwarf binary population at different age points will provide a comprehensive view of the formation mechanisms and the role of dynamics in the evolution or destruction of substellar multiple systems.  Additionally, the full characterization of the physical and orbital parameters of multiple systems will test substellar formation and evolutionary models, allowing these theories to be applied to the older population of single brown dwarfs and giant planets.

\textbf{Tracing planetary formation using D/H with ELTs.} Substellar compositional studies aid in distinguishing between planet-like accretion processes within disks, subject to disk evolution (gas freezeout into ices, dust grain chemistry), and star-like molecular cloud fragmentation, which may retain protostellar abundances. Next generation facilities will be capable of measuring molecular species in cool  atmospheres ($\sim$200-1800K) indicative of formation channel, e.g., \emph{JWST} measurements of the relative abundance of deuterium to hydrogen in the form of CH$_3$D and HDO absorption
~\citep{2018arXiv181004241M}. Observations of these species and other diagnostics (e.g., C/O ratios) with high dispersion spectroscopy will also yield abundance comparisons that can delineate between formation pathways (see white paper on high resolution RV by A. Burgasser).


\textbf{Mapping Y dwarf binary orbits with \textit{JWST}.} The unprecedented sensitivity of~\textit{JWST} at NIR and MIR wavelengths will enable high contrast observations with kernel phase interferometry (e.g.~\citep{2013ApJ...767..110P}, better than 100 at separations $<150$\,mas; Ceau et al., submitted) of the coldest and faintest brown dwarfs, achieving lower mass ratios than with previous NIR studies at $\sim1-2$\,AU. Assuming an average binary separation of $1-2\,$AU, astrometric monitoring with NIRISS full pupil imaging over the duration of the mission would map a significant portion of the orbit combining relative astrometry between the two components from imaging and astrometry of the primary relative to background sources.

\textbf{The need for an infrared~\textit{Gaia} mission.} Full orbit determination is an expensive endeavor that requires several high resolution observation epochs spanning a substantial orbital arc. As a result, only a few dozen orbits have been fully characterized. An inexpensive avenue to gather dynamical mass measurements in bulk is by calculating accelerations in the proper motions between the \textit{Hipparcos} and \textit{Gaia} epochs~\citep{2018ApJS..239...31B}. Stars common to both catalogs with discrepant proper motions are likely perturbed by a faint, hidden companion. With accelerations in the tangential direction, radial velocity measurements, and one epoch of imaging, individual dynamical masses can be measured analytically (see~\citep{2018A&A...615A.149C,2019ApJ...871L...4D} and white paper by T. Dupuy). Knowing dynamical masses will allow us to pick ideal candidates for high resolution follow up to attain full orbits. This approach is also suitable to identify long-period companions, a parameter space inaccessible until now. While this procedure is currently available only for \textit{Hipparcos} stars, future \textit{Gaia} data releases will allow a similar treatment for \textit{Gaia}-only stars with a shorter temporal baseline. However, since \textit{Gaia} was developed for optical wavelengths, it misses most brown dwarfs, which are brightest in near- and mid-infrared wavelengths~\citep{2018ApJ...862..173T}. A high-precision, near-infrared, space-based astrometric mission homologous to \textit{Gaia} would enable comparable science for cool and heavily embedded stars, allowing us to reliably model the orbits for binary systems whose orbital periods are up to twice the total mission length \citep{2008AandA...482..699C}, but can be up to several times longer. Current \textit{Gaia} internal work shows acceleration solutions can be achieved with as little as 10\% of the orbit. A new astrometric mission 10-20 years after \textit{Gaia} would therefore allow orbital fitting for brown dwarf binaries with orbital periods up to 50 years and, under the assumption that systematic focal plane differences between \textit{Gaia} and the new mission can be modeled, up to 200 years for objects seen by \textit{Gaia} (see also white paper by J. Davy Kirkpatrick).

\textbf{A library of fully characterized benchmarks.} We need to build a library of benchmark binary systems at different ages, with a variety of primary masses, mass ratios, separations, eccentricities, and configurations, observed with a combination of binary detection techniques. Ideal targets for this library are substellar eclipsing binaries. While some examples may arise from the~\textit{Kepler, K2, TESS} and future LSST missions, targeted programs like SPECULOOS~\citep{2018SPIE10700E..1ID} and MEarth~\citep{2008PASP..120..317N} should be more successful in identifying these types of systems. This venture is an expensive upfront investment, but crucial for improving model relations of luminosity, mass, and composition. 

\pagebreak
\section*{References}

\setlength{\bibsep}{-1pt plus 0.25ex}
\renewcommand\refname{\vskip -8mm}
\footnotesize{
\begin{multicols}{2}

\end{multicols}}

\end{document}